\documentclass[pre,aps,showpacs,twocolumn]{revtex4-1}

\usepackage{graphicx}
\usepackage{amsmath}
\usepackage{amsfonts}
\usepackage{amssymb}
\usepackage{natbib}
\usepackage{color}

\newcommand{\bea}{\begin{eqnarray}}
\newcommand{\eea}{\end{eqnarray}}
\newcommand{\beq}{\begin{equation}}
\newcommand{\eeq}{\end{equation}}
\newcommand{\bit}{\begin{itemize}}
\newcommand{\eit}{\end{itemize}}
\newcommand{\Y}{\mathcal{Y}}

\newcommand{\D}{\mathrm{d}}
\newcommand{\A}{\mathcal{A}}

\renewcommand{\P}{\mathcal{P}}

\newcommand{\T}{\mathcal{T}}

\begin{document}
\title {Inequalities generalizing the second law of thermodynamics for transitions between non-stationary states}
\author{Gatien Verley$^1$}
\author{Rapha\"el Ch\'etrite$^2$}
\author{David Lacoste$^1$}
\affiliation{$^1$ Laboratoire de Physico-Chimie Th\'eorique - UMR CNRS Gulliver 7083,
ESPCI, 10 rue Vauquelin, F-75231 Paris, France}
\affiliation{$^2$ Laboratoire J. A. Dieudonn\'e, UMR CNRS 6621, Universit\'e de Nice Sophia-Antipolis, Parc Valrose, 06108 Nice Cedex 02, France}
\date{\today}

\begin{abstract}

We discuss the consequences of a variant of the Hatano-Sasa relation in which a non-stationary distribution is used in place of the usual stationary one.
We first show that this non-stationary distribution is related to a difference of traffic between the direct and dual dynamics. With this formalism, we extend the definition of the adiabatic and non-adiabatic entropies introduced by M. Esposito and C. Van den Broeck in Phys. Rev. Lett. 104, 090601 (2010) for the stationary case. We also obtain interesting second-law like inequalities for transitions between non-stationary states.
\end{abstract}

\maketitle

%%%%%%%%%%%%%%%%%%%%%%%%%%%%%%%%%%%%%%%%%%%%%%%%%%%%%%%%%%%%%%%%%%%%%%%%%%%%%%%
%\subsection{Introduction}
\label{Intro}
%%%%%%%%%%%%%%%%%%%%%%%%%%%%%%%%%%%%%%%%%%%%%%%%%%%%%%%%%%%%%%%%%%%%%%%%%%%%%%%
\emph{Introduction}-
The second law of thermodynamics provides fundamental limitations on the way transitions between equilibrium states can occur.
For many years, this principle could only be expressed as an inequality.
A broad number of works summarized under the name of fluctuations theorems (FT) \cite{Evans1993_vol71,*Gallavotti1995_vol74,*Jarzynski1997_vol78,*Kurchan1998_vol31,*Lebowitz1999_vol95,*Crooks2000_vol61,Hatano2001_vol86}, have changed that fundamentally by providing equalities valid for systems arbitrarily far from equilibrium.
In particular, one equality defines the entropy production as the amount of time-symmetry breaking \cite{Maes2003_vol110a,*Gaspard2004_vol117,*Kawai2007_vol98}, a statement which not only encompasses the second law, but also sharpens it by providing additional implications at the trajectory level.
This notion of trajectories is also meaningful for stochastic optimization problems \cite{Aurell2011_vol106} and more generally for application to information theory. In this respect, the recent generalization of the Jarzynsky relation for systems operating under feedback control \cite{Sagawa2010_vol104} appears  particularly significant. With all these exciting developments, the second law of thermodynamics, a rather old idea, appears nowadays more alive than ever.

In these generalizations, an essential step was made by Hatano and Sasa \cite{Hatano2001_vol86}, who introduced the functional
 \beq
Y[c]=\int_0^T \D t \dot{h}_t \partial_h \phi(c_t,h_t),
\label{HS}
\eeq
where $\phi(c,h)=-\ln p_{st}(c,h)$, and $p_{st}(c,h)$ is the stationary probability distribution to be in a microstate
$c$ with a constant value of the control parameter $h$. They have shown that $\langle \exp(-Y[c]) \rangle=1$, a relation which has been confirmed experimentally with small colloidal systems \cite{Trepagnier2004_vol101}. Their relation implies $\langle Y \rangle \ge 0$, which translates into a modified second-law for transitions between non-equilibrium steady states (NESS) \cite{Sasa2006_vol125}, and through an expansion it also leads to a modified fluctuation-dissipation theorem for systems near a NESS \cite{Chetrite2008_vol,*Prost2009_vol103,*Seifert2010_vol89,*Verley2011_vol93}.
In this framework, in the limit of very slow driving, the probability distribution assumes at all times its stationary form evaluated at the value of the control parameter at this time. In contrast to this, we develop in this paper a framework for situations in which even in the limit of slow driving the probability distribution stays time-dependent. This extension is important to treat the following situations: (i) the system never reaches a stationary state on any reasonable time scale, as in aging systems, (ii) the system is driven by at least two control parameters, so even when the protocol $h$ is constant the dynamics remains non-stationary, or (iii) the system is prepared in a non-stationary distribution by the choice of initial conditions and is then further driven.
In \cite{Chernyak2006_vola,*Esposito2007_vol76,*Esposito2010_vol104,*Broeck2010_vol}, it was shown that remarkably, when a stationary distribution is used as reference, the second law can be split into two components, the so called adiabatic part (corresponding to the contribution of the entropy production in the limit of slow driving) and the remaining non-adiabatic part.
In this paper, we ask whether in the particular situations mentioned above, the second law can still be split into two components similar to the adiabatic and non-adiabatic entropy productions.

In order to investigate this question, we consider a system which evolves according to a continuous-time Markovian dynamics of a pure jump type. We denote a trajectory by $[c]=(c_0,c_1,...,c_N;\tau_1,.., \tau_N)$ where the $c_i$ are the states which are visited by the system and $\tau_i$ are the jumping times to go from $c_{i-1}$ to $c_i$. This trajectory is recorded from time 0 to time $T$. The transition rate to jump from a configuration $c$ to a configuration $c'$ is denoted $w_t^{h_t}(c,c')$. In this notation, the superscript $h_{t}$ refers to the dependance on the driving protocol, while the subscript $t$ refers to an additional time-dependence not related to the protocol $h$. The path probability for this trajectory is
\bea
\P[c]&=& p_0(c_0) \left[ \prod_{j=1}^{N} \exp \! \left( \! - \! \!\int_{\tau_{j-1}}^{\tau_{j}} \! \! \!\! \! \! \D t \lambda^{h_t}_t(c_{j-1}) \right) \! w^{h_{\tau_j}}_{\tau_j}(c_{j-1},c_{j}) \right]  \nonumber \\
& \times & \exp \left( - \int_{\tau_{N}}^{T} \D t \lambda_t^{h_t}(c_{N}) \right),
\eea
where $\lambda^{h_t}_t(c')=\sum_{c \neq c'} w^{h_t}_t(c',c)$ represents the escape rate to leave the state $c'$ and $p_0(c_0)$ is the probability distribution of the initial condition $c_0$.
Let us consider a logratio of path probabilities of the form:
\beq
\Delta \A[c] = \ln \frac{\P[c]}{\tilde{\P}[c^*]},
\label{def A}
\eeq
where $\tilde{\P}$ results from the application of an involution $\sim$ on the path probability $\P$, while $[c^*]=(c_0^*,c_1^*,..,c_N^*;\tau_1^*,.., \tau_N^*)$ results from a different involution acting on the trajectory $[c]$. In the following, we assume that the involution $*$ is either the identity ($[c^*]=[c]$) or the time-reversal symmetry ($[c^*]=[\bar{c}]=(c_N,c_{N-1},..,c_0;T-\tau_N,..,T-\tau_1)$).
By computing the logratio of these probabilities, one finds that
\bea
\Delta \A[c] &=& \ln \frac{p_0(c_0)}{\tilde{p_0}(c_0^*)} - \int_0^T dt [  \lambda^{h_t}_t(c_t) - \tilde{\overset{*}{\lambda}}\,^{h_t}_t(c_t) ] \nonumber \\
&+& \sum_{j=1}^N \ln \frac{w_{\tau_j}^{h_{\tau_j}}(c_{j-1},c_j)}{\tilde{w}_{\tau_j^*}^{h_{\tau_j^*}}(c_{j-1}^*,c_j^*)},
\label{explicit A}
\eea
with $c_t=c_j$ if $t \in [\tau_j,\tau_{j+1}[$ and $ \overset{*}{\lambda}\,\!^{h_t}_t = \lambda^{h_{T-t}}_{T-t}$ if the involution $*$ is the time reversal. Note that the second term in $\Delta \A[c]$ corresponds to a difference of traffic (i.e. time integrated escape rates) between the dynamics generated by $\P$ and that generated by $\tilde{\P}$, and has similarities with the dynamical activity introduced in Ref.~\cite{Maes2006_vol96,*Baiesi2009_vol103,*Maes2011_vol107}.

Now introducing $ \tilde{P}(\Delta \A)=\sum_{[c]} \delta(\Delta \A - \Delta \tilde{\A}[c]) \tilde{\P}[c]$, and using the relation $\Delta \A[c]=-\Delta \tilde{\A}[c^*]$, which follows from Eq.~\ref{def A}, one obtains a detailed fluctuation theorem (DFT) for $\Delta \A$, namely \beq
P(\Delta \A) = \exp \left( \Delta \A \right) \tilde{P}(-\Delta \A). \label{DFT}
\eeq
As a first application of Eq.~\ref{explicit A}, we choose the involution star to be the time-reversal symmetry and the involution tilde to be the time-reversal for the rates that we denote with a bar, such that $\bar{w}^{h_\tau}_\tau(c,c') = w^{h_{T-\tau}}_{T-\tau}(c,c')$. In this case, the second term in Eq.~\ref{explicit A} is zero, and $\Delta \A$ represents the total entropy production, $\Delta S_{tot}$, which satisfies a DFT of the form above   \cite{Seifert2005_vol95}.

\emph{First main result}-
We now introduce a new involution, namely the duality transformation, that we denote by a hat ($\wedge$). In analogy with the stationary case \cite{Esposito2010_vol104}, we define the dual dynamics by the following transformation of the rates:
\beq
\hat w^{h}_\tau(c,c')= \frac{w^h_\tau(c',c)\pi_\tau(c',h)}{\pi_\tau(c,h)},
\label{duality}
\eeq
where the distribution $\pi_t(c,h)$ represents the probability to observe the system in the state $c$ at a time $t>0$ in the presence of a constant (time independent) driving $h$. This distribution, which plays a key role here satisfies
\beq
\left( \frac{\partial \pi_t}{\partial t}  \right) (c,h) = \sum_{c'}  \pi_t(c',h) L_t^{h}(c',c),
\label{defpi}
\eeq
where $L_t^{h}$ is the generator defined by $L^{h}_t(c',c) = w^{h}_t(c',c) - \delta(c,c')\sum_{c''} w^{h}_t(c',c'')$. We emphasize that $\pi_t(c,h_t)$ depends only on the driving at time $t$ unlike $p_t(c,[h_t])$, the solution of the master equation with the same initial condition but with the generator $L^{h_t}_t$, which depends functionally on the driving history $[h_t]$ up to time t. Using the duality transformation introduced above, we consider the following two cases for the involutions entering in Eq.~\ref{def A}:
Case (A) where the involution $\sim$ is the combination of duality plus time reversal ($\barwedge$) and $*$ is the time reversal, with $\Delta \A[c] = \Delta A_{na}[c]$; and case (B) where the involution $\sim$ is the duality ($\wedge$) and $*$ is the identity, and $\Delta \A[c] = \Delta B_a[c]$.
In both cases, the integral in the r.h.s of Eq.~\ref{explicit A} is the same and, using Eq.~\ref{duality} and Eq.~\ref{defpi}, it can be written as
\beq
\int_{0}^{T} \D t ( \lambda^{h_t}_t(c_{t})- \hat \lambda^{h_t}_t(c_{t})) = -
\int_{0}^{T} \D t \left(\partial_t \ln \pi_t \right)(c_t,h_t).
\label{explicit T}
\eeq
We call that quantity $\Delta \T[c]$ in the following. Note that in Eq.~\ref{explicit T}, the time derivative acts only on $\ln \pi_t$, but not on the arguments of that function. This relation is our first result. It establishes an important link between the difference of traffic associated with the direct and dual dynamics and the accompanying distribution $\pi_t(c,h_t)$.

\emph{Second main result}-
We now show that using the quantity $\Delta \T$ we can generalize the notions of adiabatic and non-adiabatic contribution to the total entropy production, denoted respectively $\Delta S_a$ and $\Delta S_{na}$. We define them to be
\bea
\Delta S_{na}[c] &=&  \ln \frac{p_0(c_0)}{p_T(c_T,[h_T])}
+ \sum_{j=1}^N \ln \frac{\pi_{\tau_j}(c_j,h_{\tau_j})}
{\pi_{\tau_j}(c_{j-1},h_{\tau_j})}, \label{Sna} \\
\Delta S_a[c] &=&  \sum_{j=1}^N \ln \frac{w_{\tau_j}^{h_{\tau_j}}(c_{j-1},c_j) \pi_{\tau_j}(c_{j-1},h_{\tau_j})}
{w_{\tau_j}^{h_{\tau_j}} (c_j,c_{j-1}) \pi_{\tau_j}(c_j,h_{\tau_j})}.
\label{Sa}
\eea
where $\pi_t(c,h)$ replaces again the stationary distribution in the usual definition \cite{Esposito2010_vol104}.
These two quantities are such that $\Delta S_{tot}=\Delta S_a + \Delta S_{na}$. Since $\Delta S_{tot}$ can be further split into reservoir entropy $\Delta S_r$ and system entropy with $\Delta S=\ln {p_0(c_0)} - \ln {p_T(c_T,[h_T])}$, one can introduce an excess entropy $\Delta S_{ex}$ such that $\Delta S_r=\Delta S_a + \Delta S_{ex}$ and $\Delta S_{na}=\Delta S + \Delta S_{ex}$.
Unfortunately, the splitting into adiabatic and non-adiabatic contributions does not have in general the property that each term satisfies a DFT, although the joint distribution of $\Delta S_a$ and $\Delta S_{na}$ satisfies such a relation \cite{Garcia-Garcia2010_vol82,*Garcia-Garcia2011_vol}. Nonetheless, being of the form of Eq.~\ref{def A}, $\Delta A_{na} =  \Delta S_{na}-\Delta \T$ and $\Delta B_a = \Delta S_a - \Delta \T $ do verify separately a DFT
\beq
\ln \frac{P(\Delta A_{na})}{\hat{\bar{P}} (-\Delta A_{na})}= \Delta A_{na} \quad,
\label{DFT A B} \quad \ln \frac{P(\Delta B_{a})}{\hat{P} (-\Delta B_{a})}= \Delta B_{a}.
\eeq
These relations represent the second main result of this paper, which we now discuss in more details:

Let us assume that the driving starts at time $t_{di}>0$ and ends at time $t_{df}<T$ for a total duration $t_d=t_{df}-t_{di}$. When $\pi_t(c,h)$ relaxes very quickly to the stationary distribution (on a time scale $\tau_{st}$ such that $\tau_{st} \ll T$ and $\tau_{st} \ll t_d$), one recovers from Eq.~\ref{Sna} and Eq.~\ref{Sa} the usual definitions of the non-adiabatic and adiabatic parts of the entropy production. In this case $\Delta \T=0$, and Eq.~\ref{DFT A B} become the usual detailed fluctuation theorems satisfied by the adiabatic and non-adiabatic entropies \cite{Esposito2010_vol104}.

We notice now that $\Delta A_{na} = \Delta S_b + \Y_T$, where $\Delta S_b = \Delta S - \Delta \psi $ is a boundary term, with $\Delta \psi=\ln \pi_0(c_0,h_0) - \ln \pi_T(c_T,h_T)$ and
\beq
\Y_T[c] =  \int_0^T\D t \dot h_t \partial_h \psi_t(c_t,h_t).
\label{defY}
\eeq
This quantity is the exact analog of Eq.~\ref{HS}, when the stationary distribution $p_{st}(c,h)$ is replaced by $\pi_t(c,h)=  \exp( - \psi_t(c,h))$. Then, a consequence of Eq.~\ref{DFT A B} is that the functional $\Y_T$ satisfies a generalized Hatano-Sasa relation
\beq
\langle \exp \left( -\Y_T [c] \right) \rangle =1
\label{modified HS}
\eeq
when we consider transitions between non-stationary states, that is to say when the time $T$ is such that $T-t_d \gg \tau$ where $\tau$ is the relaxation time of $p_t(c,[h_t])$ towards $\pi_t(c,h_t)$. Since $p_0(c_0)=\pi_0(c_0,h_0)$, in this case the boundary term $\Delta S_b$ vanishes. We note that by expanding Eq.~\ref{modified HS} we have a modified fluctuation-dissipation theorem valid for systems near a general non-equilibrium state \cite{Chetrite2009_vol80,Verley2011_vol}.

Note also that the remaining parts in the entropy production $\Delta A_a=\Delta S_{tot}-\Delta A_{na}=\Delta S_a + \Delta \T$ and $\Delta B_{na}=\Delta S_{tot}-\Delta B_a = \Delta S_{na} + \Delta \T$ do not satisfy a DFT of the form of Eq.~\ref{DFT} except in the stationary case due to the fact that in this case $\Delta \T = 0$.

\emph{Third main result}-
Using the Jensen's inequality on integrated fluctuation theorem associated to Eq.\ref{DFT A B}, we get $\langle \Delta A_{na} \rangle \ge 0$ and $\langle \Delta B_a \rangle
\ge 0$. We can equivalently write that $\langle \Delta S_{na} \rangle \ge \langle \Delta \T \rangle$ and $\langle \Delta S_a \rangle \ge \langle \Delta \T \rangle$, which taken together imply $\langle \Delta S_{tot} \rangle  \ge \max (2 \langle \Delta \T \rangle, 0)$. These inequalities are very general, they hold for Markov processes in finite time $T$, arbitrary initial probability distribution $p_0$, arbitrary driving and arbitrary dynamics of the system at constant time-independent driving.
%\textcolor{blue}{The existence of lower bounds for these entropies can be understood from the fact that the dynamics controlled by $\pi_t(c,h)$ results from the original one controlled by $p_t(c)$, via a coarse-graining \cite{Kawai2007_vol98}.}
%Note that there is no lower bounds for $\Delta A_a$ or $\Delta B_{na}$, which need not be positive on average. That this should be the case can be understood by considering the particular case where the system has been prepared in a non-equilibrium state by the application of a protocol, which is then exactly compensated by the driving $h_t$. In this case, the system is in equilibrium at all times, the rates satisfy a detailed balance condition, and $\Delta S_a=-\Delta S_{na}$. Therefore, $\langle \Delta A_{a} \rangle  \le 0$ and $\langle \Delta B_{na} \rangle  \le 0$.
Note that there is no lower bounds for $\Delta A_a$ or $\Delta B_{na}$, which need not be positive on average. That this should be the case can be understood by considering a system at equilibrium on which two protocols that exactly compensate each other are applied and only the second protocol is considered as driving.
In this case, the system is in equilibrium at all times, the rates satisfy a detailed balance condition, and $\langle \Delta S_{tot} \rangle =0$. Therefore, $\langle \Delta A_{a} \rangle  \le 0$ and $\langle \Delta B_{na} \rangle  \le 0$.

From the inequality $\langle \Delta A_{na} \rangle \ge 0$, one obtains
\beq
\langle \Delta S \rangle \ge - \langle \Delta S_{ex} \rangle + \langle \Delta \T \rangle,
\label{Clausius}
\eeq
which contains the second law for transitions between equilibrium states and the modified second law for transitions between NESS \cite{Hatano2001_vol86,Sasa2006_vol125} as particular cases. For this reason, we call Eq.~\ref{Clausius} a modified second law for transitions between non-stationary states. Alternatively, one has also $\langle \Y_T \rangle \ge - \langle \Delta S_b \rangle =  D(p_T||\pi_T) \ge 0$. The equality in these relations holds in the adiabatic limit, corresponding to infinitively slow driving on $h_t$ in which case $p_T$ has relaxed towards $\pi_T$ and thus $\langle \Delta S_b \rangle  = 0$. In this limit, $\Delta A_{na}=0$ which justifies the adiabatic/non-adiabatic terminology;  $\Delta A_a=\Delta B_a=\Delta S_{tot}$ and $\Delta B_{na} = 0$. Taken together, these relations imply that $\Delta \T=0$ and that the second equation in Eq.~\ref{DFT A B} becomes the DFT satisfied by the total entropy production. Clearly, the driving can be slow even if $\pi_t(c,h_t)$ has not relaxed to a stationary distribution. Note also, that in this adiabatic limit, the Shannon entropy constructed from $\pi_t(c,h)$, namely $\langle \Delta \psi \rangle$ equals the opposite of the excess entropy $- \langle \Delta S_{ex} \rangle$ as in the case of NESS.
\begin{figure}
\begin{center}
\hspace{-0.7cm} \includegraphics[width= 9cm]{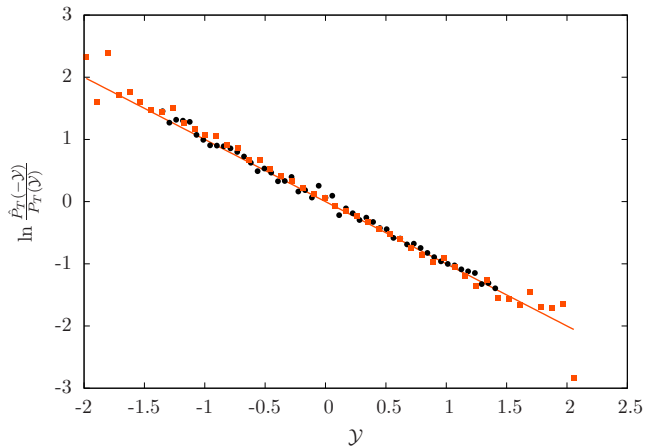}
\end{center}
\caption{Illustration of the detailed fluctuation theorem satisfied by $\Y_T$, namely Eq.~\ref{DFT A B}, in the particular case that $\Delta A_{na}=\Y_T$ and for the dynamics of a two states model. The square symbols corresponds to a fast relaxation as compared to the driving, whereas the opposite is true for the circles.}
%\caption{Detailed fluctuation relation of Eq.~\ref{DFT A_na}. The parameters for this simulation are the following: final time of the simulation %$T=15.7$ with non perturbed rates $w_0(a,b)=w_0(b,a)=1.0$ for the empty symbols ($T=5.0$ and $w_0(a,b)=w_0(b,a)=0.1$ for the full symbols), the %amplitude of the force protocol $H=3.0$, the time step of discretization for measurement of physical quantities $dt=0.005$, the force step of %discretization $H/N=3/50$, and the initial probability of state $b$ is $P_0(b)=0.9$. The probability distributions have been obtain with $10^5$ %trajectories.}
\label{fig1}
\end{figure}

\begin{figure}
\begin{center}
\hspace{-0.7cm} \includegraphics[width= 9cm]{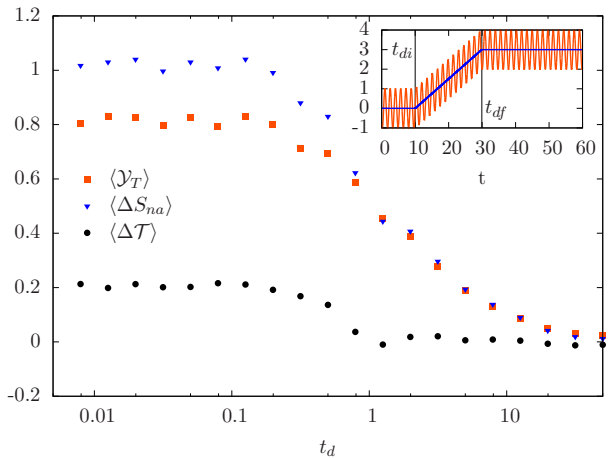}
\end{center}
\caption{Driving entropy production $\langle \Y_T \rangle$, non adiabatic entropy production $\langle \Delta S_{na} \rangle$ and $\langle \Delta \T \rangle$ as function of the duration of the driving $t_d$. The inset represents the total driving protocol, with the oscillating part at pulsation $\omega_0$ representing the reference and the solid line representing a linear protocol $h_t$ which is on between times $t_{di}$ and $t_{df}$.}
\label{fig2}
\end{figure}

\emph{Example}-
As an illustrative example, we consider a time-dependent two states model, which may be
realized experimentally in quantum optics for instance \cite{Fox2006_vol}.
We have chosen for simplicity rates of the form
$w^{h_t}(a,b)=w(a,b)e^{-h_t/2}$ and $w^{h_t}(b,a)=w(b,a)e^{h_t/2}$,
where the driving $h_t$ follows a time-symmetric half sinusoidal protocol of duration $t_d=T$.
The initial probability to be in one of the two states is chosen according to an arbitrary value different from the stationary value (here we chose arbitrarily $p_0(b)=0.9$).
Therefore, the reference dynamics is non stationary by the choice of initial condition. Through extensive kinetic Monte Carlo simulations of the trajectories followed by this system, we determine the distributions $\pi_t(c,h)$, the corresponding dual dynamics, and the distribution of $\Y_T$ from this data.
Figure \ref{fig1} shows that the first DFT of Eq.~\ref{DFT A B} is well obeyed in this case irrespective of whether the relaxation of the $\pi_t$ distribution is fast or not with respect to the driving. Figure \ref{fig2} illustrates transitions between two non-stationary states. The non-stationary states at given $h$ are created by sinusoidal reference protocols with rates $ w^h_t(a,b)=w(a,b)e^{-h-\sin \omega_0 t}$ and $ w^h_t(b,a)=w(b,a)e^{h+\sin \omega_0 t}$. The transition is produced by a piecewise linear driving protocol $h_t$ as shown in the inset of the figure. The various quantities $\langle \Y_T \rangle$, $\langle \Delta S_{na} \rangle$ and $\langle \Delta \T \rangle$ are shown as a function of the duration of the driving $t_d$. As expected, in the quasistatic limit $t_d \rightarrow \infty$, one has $\langle \Delta \T \rangle=\langle \Y_T \rangle=\langle \Delta S_{na} \rangle=0$, whereas $t_d \rightarrow 0$ corresponds to a quenched limit which is consistent with $\langle \Delta S_{na} \rangle - \langle \Delta \T \rangle = \langle \Y_T \rangle$. Note that the general evolution of $\langle \Y_T \rangle$ as function of $t_d$ is similar to that of the dissipated work in the equilibrium case \cite{Kawai2007_vol98}. 

In this particular example, we have chosen a simple dynamics for which the distribution $\pi_t(c,h)$ is analytically solvable. More generally for applications in complex systems, this distribution will not be available analytically, however if the system (or sub-system) of interest is of small size, the numerical determination of this distribution is possible through simulations \cite{Verley2011_vol}. Among the various strategies which can facilitate this numerical determination, one recent suggestion is to determine the distribution iteratively by starting from an approximate ansatz function \cite{Perez-Espigares2011_vola}.
%This promising idea was recently put forward in the stationary case %\cite{Perez-Espigares2011_vola}, but could also in principle be generalized to the %non-stationary situation discussed in this paper.

\emph{Conclusion}-
We have connected the accompanying distribution $\pi_t(c,h_t)$ introduced in \cite{Verley2011_vol} to the difference of traffic between the direct and dual dynamics. Using a non-stationary probability as reference, we have extended the notion of adiabatic and non-adiabatic contribution to the total entropy production. Unfortunately, the two new parts of the total entropy production do not verify separately a DFT as in the stationary case. Despite this, we have obtained two detailed fluctuation theorems with interesting consequences: a generalization of the Hatano-Sasa relation and second-law like inequalities for transitions between non-stationary states. These results could have important applications in particular in force measurements with biopolymers or proteins, for the characterization of small glassy systems, or for stochastic optimization problems.

\bibliography{Ma_base_de_papier}

\end{document}